\documentclass[aps,pra,twocolumn]{revtex4-2}
\usepackage{placeins}
\usepackage{microtype}
\usepackage{etoolbox}

\usepackage{amsmath}
\usepackage{graphicx}
\usepackage{dcolumn}
\usepackage{bm}
\usepackage{xcolor}
\usepackage{appendix}
\usepackage[mathlines]{lineno}
\newcolumntype{d}[1]{D{.}{.}{#1}}

\begin{document}

\title{Spectroscopy of the \(\mathbf{X^2\Sigma^+(v=2) \rightarrow A^2\Pi_{1/2}(v=1)}\) Transition in MgF: Hyperfine Structures and Spectroscopic Constants}

\author{Youngju Cho}
\author{Yongwoong Lee}
\author{Kikyeong Kwon}
\author{Seunghwan Roh}
\author{Giseok Lee}
\altaffiliation{Current affiliation: Department of Physics, Harvard University, Cambridge, Massachusetts 02138, USA}
\author{Eunmi Chae}
    \email{echae@korea.ac.kr}
\affiliation{Department of Physics, Korea University, Seoul, 02841, Republic of Korea}

\date{\today}

\begin{abstract}
We report spectroscopic results of the \(X^2\Sigma^+(v=2) \rightarrow A^2\Pi_{1/2}(v=1)\) transition in magnesium monofluoride (MgF). Using Doppler-free Laser-Induced Fluorescence (LIF) spectroscopy on the \(X^2\Sigma^+(v=2) \rightarrow A^2\Pi_{1/2}(v=1)\) transition, we resolved 47 hyperfine components distributed over 11 transition lines in X and A states. An effective Hamiltonian—comprising contributions from vibrational, rotational, \(\Lambda\)-doubling, and hyperfine interactions—was presented to model the energy structure of the  \(A^2\Pi_{1/2}(v=1)\) state. The spectroscopic parameters, including the rotational constant, the  \(\Lambda\)-doubling parameter, and the hyperfine interaction constants, were extracted using a least-square fitting and Markov Chain Monte Carlo (MCMC) procedure. Our study reveals that the spectroscopic constants show subtle changes compared to the  \(A^2\Pi_{1/2}(v=0)\) state. These results provide critical spectroscopic benchmarks for optimizing optical cycling schemes in MgF, thereby advancing optical cycling efficiency in the magneto-optical trapping of MgF.
\end{abstract}

\maketitle

\section{Introduction}
Diatomic molecules have emerged as particularly promising tools for quantum computing \cite{quantumcomputing1, quantumcomputing2}, quantum simulation \cite{quantumsimul1, quantumsimul2} and precision measurements \cite{precise1, precise2}. This potential arise from their complex internal energy structure, which offers a rich manifold for encoding quantum information and sensing subtle fields. To harness these features for the advancement and application of quantum science, it is imperative to establish an ultracold environment, where thermal motion is suppressed, thereby preserving fragile quantum coherence \cite{ultracoldenv1,ultracoldenv2,ultracoldenv3}.
One route to make ultracold diatomic molecular samples is direct laser cooling, a technique previously developed for neutral atoms \cite{atom1,atom2,atom3} and recently demonstrated for diatomic molecules and even for polyatomic molecules, such as SrF, CaF, BaF, YbF, AlF, YO, CaOH and SrOH \cite{SrF,CaF,BaF,YbF,AlF,YO,CaOH,SrOH}. However, the complex rovibrational structure of these molecules \cite{rovibrational0,rovibrational1,rovibrational2,rovibrational3,rovibrational4,rovibrational5,rovibrational6,rovibrational7,rovibrational8,rovibrational9,rovibrational10,rovibrational11,rovibrational12,rovibrational13,rovibrational14} presents significant technical challenges for laser cooling.

While the rovibrational structure offers a rich manifold of utilizable internal states \cite{rich1,rich2}, it simultaneously introduces additional decay channels that complicate optical cycling \cite{decay1}. As rotational transitions follow selection rules, careful choices can minimize unwanted decay \cite{MgFcooling}. However, vibrational transitions are governed by Franck-Condon factors (FCFs), making multiple repump lasers necessary to address the relevant vibrational states within the $X^2\Sigma^+$ ground state in order to form quasi-closed cycling transitions \cite{quasi1,quasi2,MgFcooling,decay1,MgFcooling6}.

For quasi-closed optical cycling in diatomic molecules, a widely used strategy is to drive transitions such as the $^2\Sigma^+\rightarrow\,^2\Pi_{1/2}$ $P_1(1)/Q_{12}(1)$ line, which serves as the main cycling transition, together with an additional $^2\Sigma^+ \rightarrow\,^2\Sigma^+$ $P_1(1)$ repumping line (see Figure 1). In the case of MgF, the main transition is the $X^2\Sigma^+(v=0) \rightarrow A^2\Pi_{1/2}(v=0)$ $P_1(1)/Q_{12}(1)$. The first repump is the $P_1(1)$ transition of $X^2\Sigma^+(v=1) \rightarrow B^2\Sigma^+(v=0)$ $P_1(1)$, and the second repump is the $P_1(1)/Q_{12}(1)$ transition of $X^2\Sigma^+(v=2) \rightarrow A^2\Pi_{1/2}(v=1)$ \cite{MgFcooling,MgFcooling2,MgFcooling3,MgFcooling4,MgFcooling5}.

\begin{figure}[t!]
\includegraphics[width=9cm]{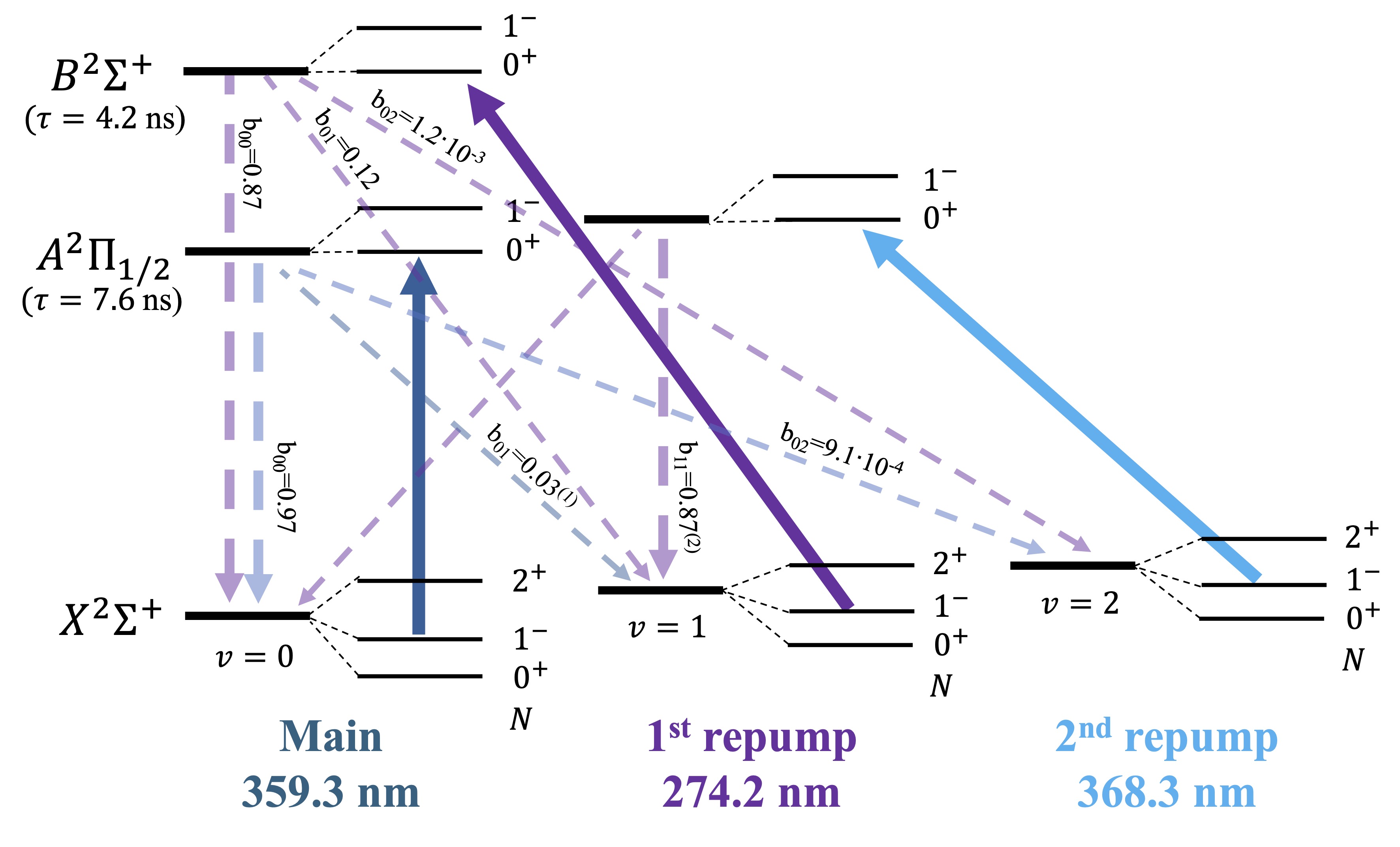}
\caption{\label{fig1}Schematic diagram of the electronic-rovibrational states and transitions for laser cooling of MgF molecules. The $X^2\Sigma^+(v=0) \rightarrow A^2\Pi_{1/2}(v=0)$ transition provides the main optical cycling path, while additional lasers at 274.2 nm and 368.4 nm repump population leaked into vibrationally excited states. The relative branching ratios $b_{v^{\prime\prime} v^{\prime}}$ and rotational quantum number $N$ are also indicated.\cite{MgFcooling, MgFcooling6}}
\end{figure}

Accurate knowledge of each transition frequency is essential for understanding the corresponding energy levels and parameters and for achieving successful laser cooling of MgF. Spectroscopic studies have been conducted on the main and first repump transitions, and recent results report their transition frequencies with an accuracy of approximately 10 MHz \cite{MgFcooling4, MgFcooling5}. However, the second repump transition has only been reported with an accuracy of about 550 MHz \cite{MgFcooling4}. In this study, we present hyperfine-resolved spectra of the $X^2\Sigma^+(v=2) \rightarrow A^2\Pi_{1/2}(v=1)$ (abbreviated as \(X(2) \to A(1)\)) transition. Also, based on these measurements, we derive the spectroscopic parameters for the $A_2\Pi_{1/2}(v=1)$ state.

\section{Theory}
\begin{figure}[hbt!]
\includegraphics[width=9cm]{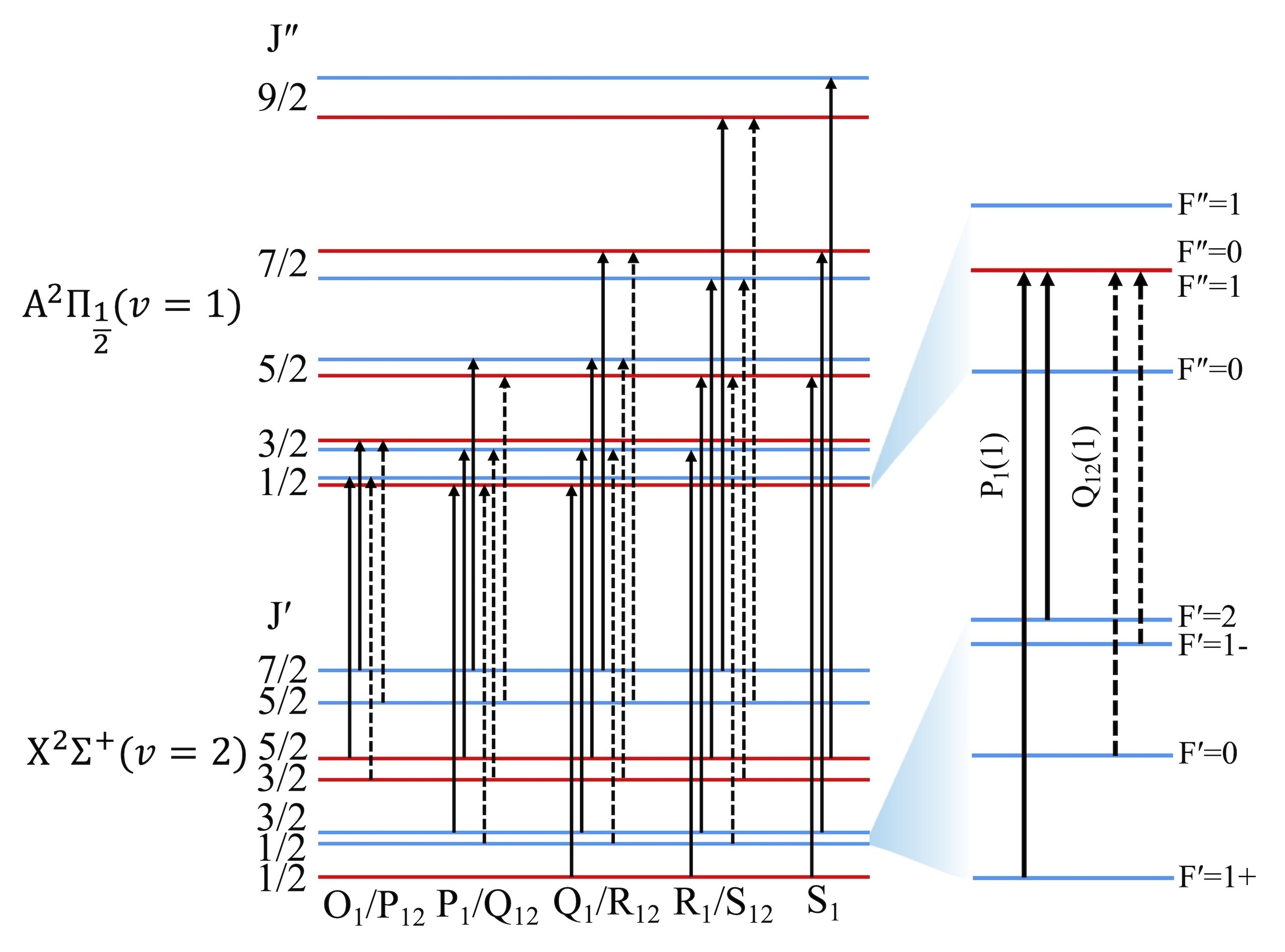}
\caption{\label{fig2} A diagram of the transition lines is provided. All possible transition lines within these energy levels are illustrated, with the \(O_1,\, P_1,\, Q_1,\, R_1,\, S_1\) drawn as a solid line and the \(P_{12},\, Q_{12},\, R_{12},\, S_{12}\) as a dashed line. The energy level’s parity is indicated by color. Red denotes the ($+$) parity state, while blue denotes the ($-$) parity state. On the right side of the figure, the hyperfine structures of the \(X^2\Sigma^+(v=2, N=1)\) and \(A^2\Pi_{1/2}(v=1, J=1/2)\) states are given and \(P_1(1)/Q_{12}(1)\) the transition lines used as the 2nd repump transition for laser cooling are drawn. Energies are not to scale for clarity.} 
\end{figure}

The energy structure and overall transition lines of MgF are illustrated in Figure \ref{fig2}. The well-defined rotational quantum numbers differ between the \(X(2)\) and \(A(1)\) state because they belong to different Hund's coupling cases \cite{rovibrational7,rovibrational12}. The \(X\) state follows Hund’s case (b), where the weak spin-orbit coupling makes \(N\) a good quantum number, and the quantum number \(J\) follows \(J=|N\pm S|\). In contrast, the \(A\) state follows Hund’s case (a), where strong spin-orbit coupling results in \(J\) being well-defined. 
As the spectroscopic constants for the $X^2\Sigma^+(v=2)$ state have already been measured with high accuracy \cite{MgFX2_1,MgFX2_2,MgFX2_3}, the present work is focused on the analysis of the \(A(1)\) state. An effective Hamiltonian to describe the \(A(1)\) state of MgF is constructed as a sum of several energy contributions: vibronic energy, rotational energy, \(\Lambda\)-doubling effects, and hyperfine interactions following Hund's case (a) \cite{rovibrational12}. It is expressed as
\begin{equation}
  H_{\text{eff}} = T_{v=1} + H_{\text{rot}} + H_{\Lambda} + H_{\text{hf}},
  \label{eq1}
\end{equation}
where \(T_{v=1}\) represents the origin of the \(X(2) \rightarrow A(1) \) transition, which includes the electronic and vibrational energy differences between the two states. The rotational part of the Hamiltonian, denoted as \(H_{\text{rot}}\), is modeled under the rigid rotor approximation and takes the form
\begin{equation}
  H_{\text{rot}} = B\,\mathbf{J}^2 - D\,\mathbf{J}^4.
  \label{eq2}
\end{equation}
Here \(B,\, D\) is the rotational constant. \\
\(\Lambda\)-doubling in diatomic molecules arises from the interaction between nuclear rotation and electronic orbital angular momentum. It can be written in the form of
\begin{equation}
  H_{\Lambda} = 
  \mp\frac{1}{2}(-1)^{J-1/2}(p+2q) \left(J+\frac{1}{2} \right)
  \label{eq3}
\end{equation}. The minus (plus) sign at the very beginning of the expression applies to the positive (negative) parity state. \(\Lambda\)-doubling affects the small energy splitting observed between levels of opposite parity with the same J and can change subtly depending on different vibrational states due to the alteration in the electronic charge distribution \cite{lambdadoubling1, lambdadoubling2}.

Hyperfine interactions, which stem from the coupling between the nuclear spin \(\mathbf{I}\) and the electronic angular momenta, are described by the following term:
\begin{equation}
\begin{split}
  H_{\text{hf}} =\; &a\,I_zL_z + b_F\,\mathbf{I}\cdot\mathbf{S} + \frac{c}{3}\,\Bigl[ 3(\mathbf{I}\cdot \mathbf{n})(\mathbf{S}\cdot \mathbf{n}) - \mathbf{I}\cdot \mathbf{S} \Bigr]\\
  &- \frac{d}{2}(S_+I_++S_-I_-)
\end{split}
\label{eq4}
\end{equation}
In this formula, the parameter \(a\) represents the nuclear spin-orbit coupling, which is the isotropic coupling between the nuclear spin and the electron spin density at the nucleus. The constant \(b_F\) describes the Fermi-contact term, and \(c\) accounts for the anisotropic portion of the electron-nuclear dipolar interaction. The term involving \(d\) represents dipole-dipole coupling. The vector \(\mathbf{n}\) is a unit vector directed along the internuclear axis. By summing these contributions, the effective Hamiltonian for the \(A(1)\) state of MgF is given by
\begin{equation}
\begin{split}
  H_{\text{eff}} =\; &T_{v=1} + B\,\mathbf{J}^2 -D\,\mathbf{J}^4\mp(-1)^{J-1/2}\frac{1}{2}(p+2q) \left(J+\frac{1}{2} \right)\\
  & +a\,I_zL_z + b_F\,\mathbf{I}\cdot\mathbf{S} + \frac{c}{3}\,\Bigl[ 3(\mathbf{I}\cdot \mathbf{n})(\mathbf{S}\cdot \mathbf{n}) - \mathbf{I}\cdot \mathbf{S} \Bigr]\\
  &- \frac{d}{2}(S_+I_++S_-I_-)
\end{split}
\label{eq5}
\end{equation}
This theoretical framework creates a direct connection between the spectral observations in the \(X(2) \to A(1)\) transition and the intrinsic molecular parameters, thereby enabling the extraction of accurate values for \(T_{v=1}\), \(B\), \(a\), \(b\), \(c\), \(d\), and \(p+2q\). 
\begin{figure}[hbt!]
\includegraphics[width=10cm]{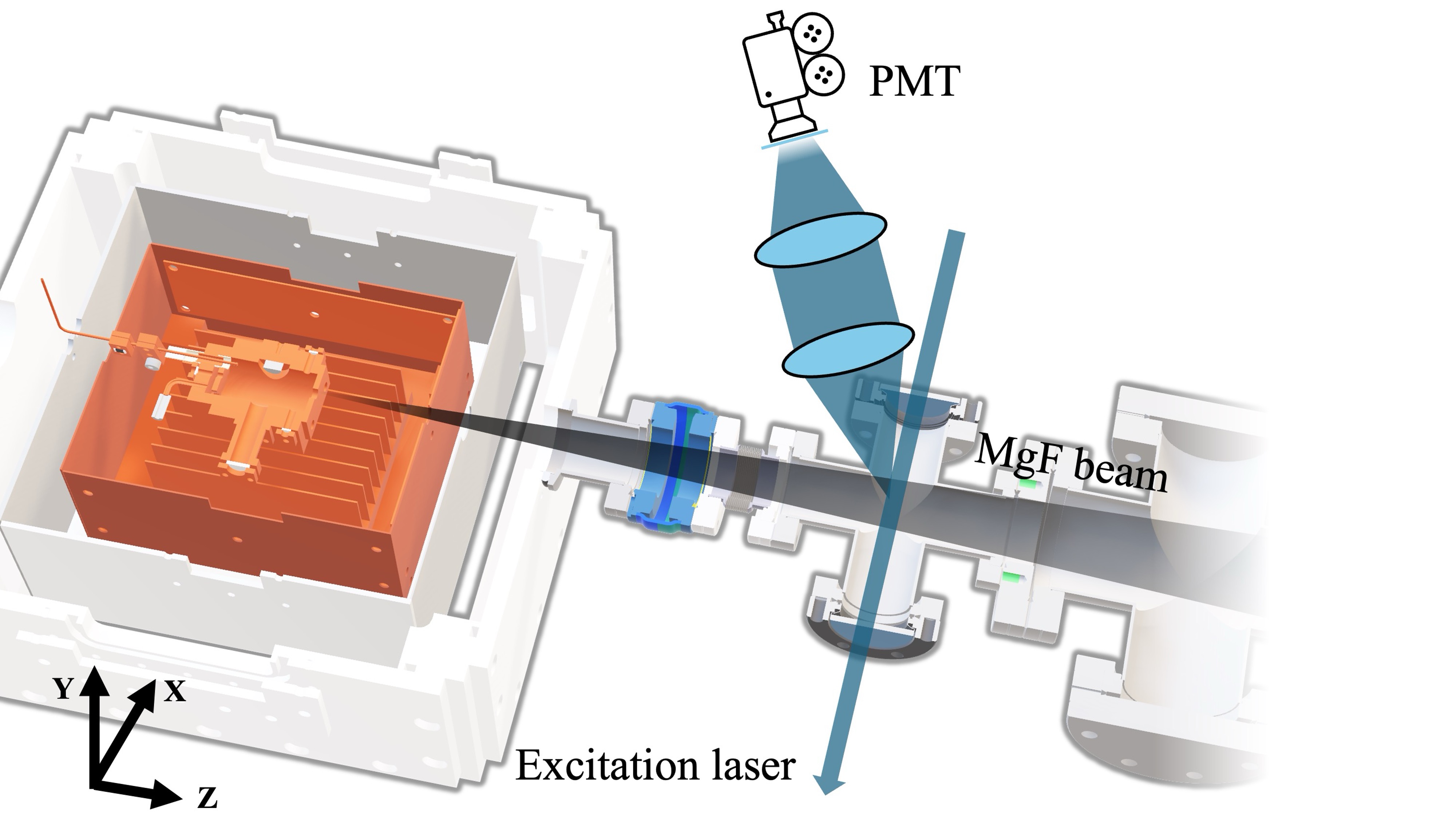}
\caption{\label{fig3}
Schematic of the experimental setup. The direction of the excitation laser is perpendicular to the MgF molecular beam propagation, and the laser-induced fluorescence (LIF) is collected by a photomultiplier tube (PMT) with an optical filter and imaging lens setup. The coordinate axes indicate that the excitation laser propagates along the $x$-axis while the MgF beam travels along the $z$-axis.
}
\end{figure}
\section{Method}
\subsection{Cryogenic Setup}
The MgF buffer-gas beam was produced using the cryogenic buffer-gas beam (CBGB) method \cite{CBGB,CBGB2}. MgF was generated by ablating an Mg target with an Nd:YAG laser with 18 mJ output in \(\mathrm{SF_6}\) environment, which act as a fluorine donor. The temperature of the cryogenic cell, where Mg attached, was maintained at approximately 4 K throughout the experiment. The temperature of the MgF gas within the main cell was determined by absorption spectroscopy. Analysis of the Doppler broadening yielded an estimated temperature range of 4-6 K. The forward velocity of the molecular beam was measured to be 180 m/s along z direction, and the transverse velocity width was approximately 4 m/s.

\subsection{Laser System}
A homemade external cavity diode laser (ECDL) is constructed based on a Littrow configuration \cite{ECDL} with a 370 nm diode. We tune the diode to 368.3 nm by cooling it to 270 K inside a hermetically sealed enclosure. We lock the ECDL frequency using a wavemeter (HighFinesse WS8-10) calibrated to the rubidium D2 line \cite{rbd2}. The absolute accuracy of the wavemeter is given as 20 MHz with calibration and the measurement resolution as 200 kHz. The ECDL frequency is stabilized by a piezometer-based proportional-integral-derivative (PID) loop, providing a frequency stability of 1.3 MHz. Given this level of stability, together with the wavemeter uncertainty, we can obtain reliable spectral frequency measurements with 20 MHz uncertainty. The frequency-tuned laser is transmitted along the \(\hat{x}\) direction of the chamber in the form of a well-defined Gaussian beam after a single-mode optical fiber. The laser has a collimated \(1/e^2\) diameter of approximately 10 mm to maximize spatial overlap to the molecular beam. The power of the laser is set to 1 mW to suppress power broadening. We obtained a Doppler-free LIF signal by directing the excitation laser perpendicularly to the molecular beam. The LIF signal is acquired by a photomultiplier tube aligned to \(\hat{y}\) the direction of the chamber with a proper UV bandpass filter to reduce background scattering.

\section{Result}

\begin{figure*}[hbt!]
\includegraphics[width=17.5cm]{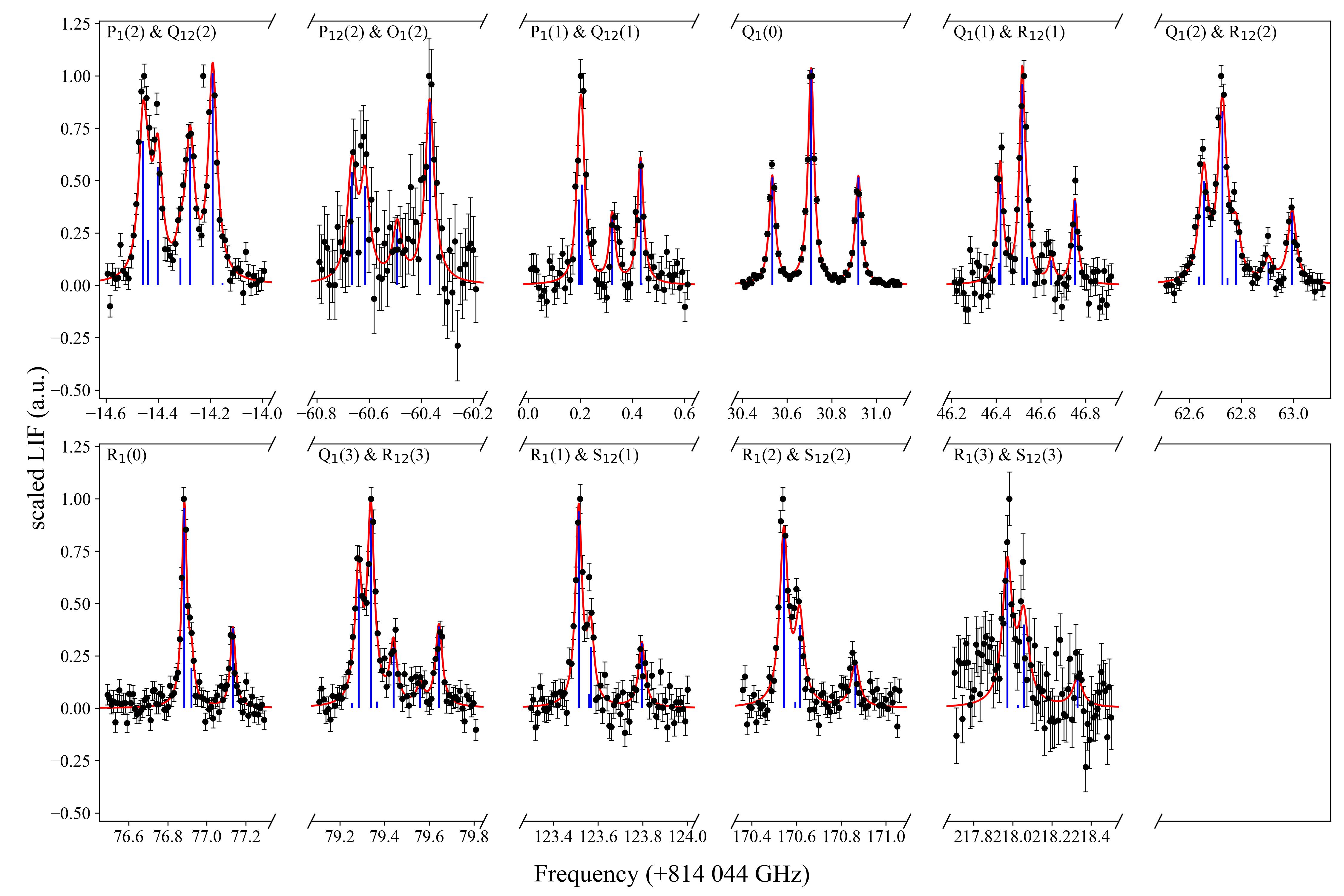}
\caption{\label{fig4} The total LIF spectra obtained in the experiment. The red lines indicate the fitting lines, and the blue vertical lines denote frequency and relative strength of each transition determined by fitting the data. The FWHM of the spectra was approximately 30-40 MHz. The reference point of the horizontal axis is at 814044 GHz. As the rotational quantum number \(N\) increases, the LIF signal tends to gradually decrease, resulting in a low signal-to-noise ratio. This can be explained by the fact that our molecular beam has a rotational temperature of a few Kelvin so that more molecules are populated in the low rotational states.}
\end{figure*}

We observed 11 transition lines in the MgF \(X(2) \to A(1)\) transition, from \(N'=0\) to \(3\) in the \(X(2)\) state to \(J''=1/2\) to \(9/2\) in the \(A(1)\) state. All the detected LIF spectra are presented in Figure \ref{fig4}, and the measured absolute frequencies for each transition line are summarized in Table 3. In total, 47 hyperfine-resolved transitions are identified. 

Initially, the transition frequencies were guessed by diagonalizing the effective Hamiltonian, which is given in Eq. \ref{eq5}. The effective Hamiltonian was computed using the PyLCP library \cite{PyLCP} in \(\textsc{python}\). The relative peak heights were obtained by calculating the absolute square of the corresponding electric dipole matrix elements. In the next step, each spectrum was fit to a Lorentzian profile to extract the Full Width at Half Maximum (FWHM) of each spectrum. At this stage, the relative peak heights were fixed to the calculated values and the frequencies were treated as the fitting parameters. In the last step, all LIF spectra were fit using molecular parameters and overall amplitudes for each spectrum, while the relative peak heights and FWHM values were fixed.

We used the Markov Chain Monte Carlo (MCMC) \cite{mcmc} and least-square algorithm \cite{leastsquare} to optimize the parameters included in the effective Hamiltonian (see Eq. \ref{eq5}). The rotational constants (\(B,D\)), hyperfine interaction constants (\(a, b_F+\frac{2}{3}c, d\)), and \(\Lambda-\)doubling constant (\(p+2q\)) are determined as summarized in Table~\ref{ta1}. Additionally, the origin of the vibronic transition, \(T_{v=1}\), was determined to be \(814063145.2 (3)\ \mathrm{MHz}\). The rotational constant (\(B, D\)) obtained in this work slightly deviates from the previously reported value \cite{MgFrot1,MgFrot2} and falls outside its quoted uncertainty range, which is attributed to the higher spectral frequency resolution achieved in our measurements.
We also confirm that the hyperfine constants show values similar to the \(A(0)\) state within the error range.
\begin{table}[t]
\caption{\label{ta1}
Fitted parameters of the \(A(1)\) state in MgF. The values in parentheses are standard errors. The \(A(0)\) state is shown for comparison. All values are in MHz.
}
\begin{ruledtabular}
\begin{tabular}{cccc}
\textrm{Parameter}&
\textrm{This work}&
\textrm{Previous work\cite{MgFrot1}}&
\textrm{$A(0)$\cite{MgFcooling2}}
\\
\colrule
\textrm{$B$}      & 15419.0(4) & 15474(2)   & 15788.2(3)    \\ 
\textrm{$D$}      & 0.023(1) &  0.023(1)  & -             \\
\textrm{$a$}        & 107(6)   & -          & 109(6)        \\
\textrm{$b_F+2c/3$}  & -52(12)  & -          &-52(14)        \\
\textrm{$d$}        & 130(1)   & -          &135(7)         \\
\textrm{$p+2q$}     & 16.5(2)  & -          &15(2)          \\
\end{tabular}
\end{ruledtabular}
\end{table}

\begin{table}[t]
\caption{\label{ta2}
The absolute frequency of the 2nd repump transition of MgF for laser cooling and a few higher transition lines. The mean and standard deviation values are given. The unit of all values is MHz.
}
\begin{ruledtabular}
\begin{tabular}{ccc}
\textrm{Transition}&
\textrm{This work}&
\textrm{Previous work\cite{MgFcooling}}
\\
\colrule
\textrm{$P_1(1)/Q_{12}(1)$}      & 814044323(20) & 814044490(550)  \\ 
\textrm{$Q_1(1)/R_{12}(1)$}      & 814090520(20) & 814090660(550)  \\
\textrm{$R_1(1)$}                & 814167510(20) & 814167600(550)  \\
\end{tabular}
\end{ruledtabular}
\end{table}

\section{conclusion}
In this study, we performed hyperfine structure-resolved spectroscopy on the \(X(2)\rightarrow A(1)\) transition of MgF, observing 11 transition lines and a total of 47 hyperfine components. With an effective Hamiltonian combined with a Markov Chain Monte Carlo and least-square fitting procedure, we successfully extracted molecular parameters including the rotational constant, \(\Lambda\)-doubling, and hyperfine constants of the A(1) state.

We also accurately determined the center frequencies of the second repump transitions of MgF for laser cooling. Although these values are approximately 170 MHz lower than the previously reported results (see Table~\ref{ta2}), our measurements remain consistent within their error bounds. Our results, with improved accuracy, contribute to a better understanding of the energy structure of the MgF molecule and will be highly beneficial for future laser slowing and MOT experiments.

\begin{acknowledgments}
Special appreciation is given to Hyeonjoon Jang and Donkyu Lim, whose expert advice on statistical analysis was indispensable in shaping the research methodology and interpreting the results. The authors acknowledge support from the National Research Foundation of Korea under grant numbers RS-2022-
NR119745, RS-2024-00439981, RS-2024-00431938, and
RS-2023-NR068116. 
\end{acknowledgments}

\begin{table*}[b]
\caption{\label{table3}
Observed transition frequencies in the $X(2)\rightarrow A(1)$ transition of MgF. 
The differences between the calculated transition frequencies from the effective Hamiltonian with the fitted parameters and the measured frequencies are listed in the rightmost column of the table. The fact that most of these differences are significantly smaller than the experimental uncertainty of 20 MHz indicates that our fitting procedure performs well. The uncertainty in these transition frequencies, 20 MHz, is determined by the absolute accuracy of the wavemeter. For the \(R_1(3)/S_{12}(3)\) transition, a single deviation exceeds the 20 MHz instrumental uncertainty. This is likely due to the low population in the X state with N=3, which resulted in a poor signal-to-noise ratio and a larger statistical uncertainty in the fit.}
\begin{ruledtabular}
\begin{tabular}{l d{9.1} d{1} d{1.1} d{1} d{1.1} d{1} d{2.1}}
\multicolumn{1}{c}{Branch} & 
\multicolumn{1}{c}{Frequency (MHz)} & 
\multicolumn{1}{c}{$N^\prime$}&
\multicolumn{1}{c}{$J^\prime$}&
\multicolumn{1}{c}{$F^\prime$}&
\multicolumn{1}{c}{$J^{\prime\prime}$}&
\multicolumn{1}{c}{$F^{\prime\prime}$}&
\multicolumn{1}{c}{Obs.-Calc.}\\
\hline
\textrm{$Q_1(0)$}           & 814074533 & 0 &  0.5&   1&  0.5&  0& 0\\
                            & 814074708 & 0 &  0.5&   1&  0.5&  1& 0\\
                            & 814074920 & 0 &  0.5&   0&  0.5&  1& 0\\                       
\textrm{$R_{1}(0)$}         & 814120883 & 0 &  0.5&   1&  1.5&  2& 0\\
                            & 814120913 & 0 &  0.5&   1&  1.5&  1& -5\\
                            & 814121129 & 0 &  0.5&   0&  1.5&  2& -1\\    
                            
\textrm{$P_1(1)/Q_{12}(1)$} & 814044196 & 1 &  1.5&   2&  0.5&  1& 0\\
                            & 814044203 & 1 &  0.5&   1&  0.5&  1& 0\\
                            & 814044323 & 1 &  0.5&   0&  0.5&  1& 0\\
                            & 814044431 & 1 &  1.5&   1&  0.5&  1& 0\\
                            
\textrm{$Q_1(1)/R_{12}(1)$} & 814090408 & 1 &  1.5&   2&  1.5&  1& -2\\
                            & 814090419 & 1 &  0.5&   1&  1.5&  1& 1\\
                            & 814090520 & 1 &  1.5&   2&  1.5&  2& 5\\
                            & 814090529 & 1 &  0.5&   1&  1.5&  2& 6\\
                            & 814090537 & 1 &  0.5&   0&  1.5&  1& 0\\
                            & 814090627 & 1 &  1.5&   1&  1.5&  1& -18\\
                            & 814090753 & 1 &  1.5&   1&  1.5&  2& 2\\
                            
\textrm{$R_1(1)/S_{12}(1)$} & 814167510 & 1 &  1.5&   2&  2.5&  2& 0\\
                            & 814167555 & 1 &  1.5&   2&  2.5&  3& 0\\
                            & 814167563 & 1 &  0.5&   1&  2.5&  2& 0\\
                            & 814167791 & 1 &  1.5&   1&  2.5&  3& 1\\

\textrm{$O_1(2)/P_{12}(2)$} & 813983340 & 2 &  1.5&   1&  0.5&  0& 6\\
                            & 813983382 & 2 &  1.5&   2&  0.5&  1& -3\\
                            & 813983516 & 2 &  1.5&   1&  0.5&  1& 7\\
                            & 813983634 & 2 &  2.5&   2&  0.5&  1& 0\\
                            
\textrm{$P_1(2)/Q_{12}(2)$} & 814029543 & 2 &  2.5&   3&  1.5&  2& 2\\
                            & 814029545 & 2 &  1.5&   2&  1.5&  2& -16\\
                            & 814029591 & 2 &  1.5&   2&  1.5&  1& -4\\
                            & 814029686 & 2 &  1.5&   1&  1.5&  2& 1\\
                            & 814029721 & 2 &  1.5&   1&  1.5&  1& 2\\
                            & 814029808 & 2 &  2.5&   2&  1.5&  2& -1\\
                            
\textrm{$Q_1(2)/R_{12}(2)$} & 814106650 & 2 &  1.5&   2&  2.5&  2& -4\\
                            & 814106723 & 2 &  2.5&   3&  2.5&  2& -1\\
                            & 814106777 & 2 &  1.5&   1&  2.5&  2& -1\\
                            & 814106902 & 2 &  2.5&   2&  2.5&  2& 0\\
                            & 814106993 & 2 &  2.5&   2&  2.5&  3& 1\\
                            
\textrm{$R_1(2)/S_{12}(2)$} & 814214541 & 2 &  2.5&   3&  3.5&  3& 1\\
                            & 814214602 & 2 &  1.5&   2&  3.5&  3& -7\\
                            & 814214847 & 2 &  2.5&   2&  3.5&  3& -10\\
                        
\textrm{$Q_1(3)/R_{12}(3)$} & 814123284 & 3 &  2.5&   3&  3.5&  3& -1\\
                            & 814123342 & 3 &  3.5&   4&  3.5&  4& 3\\
                            & 814123444 & 3 &  2.5&   2&  3.5&  3& 3\\
                            & 814123551 & 3 &  3.5&   3&  3.5&  3& -9\\
                            & 814123647 & 3 &  3.5&   3&  3.5&  4& 4\\
                    
\textrm{$R_1(3)/S_{12}(3)$} & 814261973 & 3 &  3.5&   4&  4.5&  5& -2\\
                            & 814262049 & 3 &  2.5&   3&  4.5&  4& -7\\
                            & 814262311 & 3 &  3.5&   3&  4.5&  4& -21\\
\end{tabular}
\end{ruledtabular}
\end{table*}

\bibliography{MgF.bib}
\end{document}